\documentclass[english]{article}
\usepackage[T1]{fontenc}
\usepackage[latin9]{inputenc}
\usepackage{geometry}
\usepackage{float}
\usepackage{amsthm}
\usepackage{amsmath}
\usepackage{graphicx}

\makeatletter
\providecommand{\tabularnewline}{\\}

\numberwithin{equation}{section}
\numberwithin{figure}{section}

\usepackage{chngcntr}
\counterwithout{figure}{section}
\counterwithout{equation}{section}

\makeatother

\usepackage{babel}
\begin{document}

\title{Words as block of notes and Zipf law in music using visibility algorithm.}

\author{Miguel Sanchez Islas.}
\maketitle
\begin{abstract}
In this work we were aimed to study music using visibility algorithms.
This algorithms are used as a way to get a network, from a time series.
However the main idea of this paper was not to study music from the
perspective of complex networks, but to studying it as a lenguage,
basically the changes in wath we call \textit{words,} between diferents
ages in music. The visibility algorithm provide us a way to cut time
series and generate the words, of the composers we where studying.
We study clasical music composers from differents ages, from Bach
to Schostakovish. Time series in this work came from MIDI archives
and the series were generated by, the absolute diference between two
adyacent notes, in the first voice in the MIDI archive. Once we had
our series we apply the algorithm to generate words. The visibility
algorihm was used also, as a way to get Zipf law becouse at first
what we found was that when we used visibility algorithm we create
a free-scale network, so we decided to take subgroup of this network
that is going to be also free of scale, and make the musical words. 
\end{abstract}

\section*{Introduction .}

Music has been studied in several ways long ago. One of the first
studies trying to see music behavior were made by , Voss and Clarke\cite{voss19781}.
They observe a 1/f behaivor in music pitch. Since then there exist
a number of works trying to study music, some from times series mesuring
hurst exponent\cite{voss1989random,su2006multifractal,su2007music},
treating music like a fractional Brownian motion (fBm), seeing and
searching fractality and a lot of research related . Recently a new
perspective has been used, transforming music into a complex networks
in order to studying music, from the tools of complex networks\cite{liu2009composing,liu2010complex}.
In contrast , in this work we try to see music in a different way,
we are interested in see if music can be treated as a lenguage, from
the point of view of a lenguage we need to define what our words are
going to be. In order to form what we are going to call a word, we
used the so called visibility algorithm\cite{lacasa2008time}. Next
what we are concerned, is to seeing if this words follow the Zipf's
law, like the rest of lenguages.

\subsection*{Visibility Algorithm}

One algorithm that transform a time series into a network, is the
Visibility Algorithms \cite{lacasa2008time,luque2009horizontal,nunez2012visibility}.
The natural visivility algorithm is a really easy way to transform
a time series into a network. The main idea of the algorithm, is comparing
two diferent points (A,B) in the time series. In order to decide,
if this two are going to be conected in the network, wath visibility
does is to compare the straight line that can be draw between this
points and then check if all the points that lie between (A,B) are
below the straight line, if does, then the algorithm says that the
points A and B are conected in the network.

The Visibility Algorithm, is often written as the following inequality: 

\begin{equation}
y_{c}<y_{b}+\left(y_{a}-y_{b}\right)\frac{t_{b}-t_{c}}{t_{b}-t_{a}}
\end{equation}

As we can see in figure \ref{vix}. 

Recently it has been shown that the visiibility algorithm is a way
to find motifs in time series \cite{iacovacci2016sequential}.

\subsection*{Zipff Law.}

In 1932 George Kingsley Zipf discover a new law concerning, the statistics
properties of lenguages\cite{zipf1932selected}. From this point it
has been seen that humans lenguage follow the distribit

The zipff law was discoer in ....

\section*{Method.}

In order to analyze music from diferent composers, we use MIDI files.
In this files, we have information about duration and the note. Instead
of having this information pictorically, as is often seeing in music
scores, in our case is displayed with numerical values. So its relatively
easy to extract the information and generate time series. Also we
have the information related to the voices, all the pieces analyzed
correspondo to piano works, and are from diferent composers.

Is good to clarify that, we only use the first voice in this study.
Some times a chord comes out in the voice we are analyzing, in wich
case we only used the higest note. Having extracted the information
, its easy to made a time series from adyacent notes. In this work
we were no worry about the duration of notes. We put every note the
same duration in the time series.

\begin{figure}[H]
\begin{centering}
\includegraphics[scale=0.5]{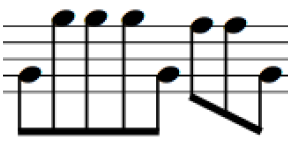}
\par\end{centering}

\caption{{\footnotesize{In this picture we can see the notes, as they are in
the pentagram . In this work we aren't worry in time series of notes,
instead we introduce a new time series that is going to be formed
by the difference in semitones between adjacent pair of notes.}}}

\label{pent}
\end{figure}

What we see in the picture above, is the representation of notes as
they are in a pentagram, normally one way of doing a time series of
music notes, is to put the values of the notes acording to their corresponding
value of the midi coding. Nevertheless, in this study we were aimed
to do the semitone diference between adjacent notes. We can think
that we are studying the discrete derivate of the original time series,
becouse the diference in time is always one. So what we have pictorically
in figure \ref{pent}, once the diference in semitones between adjacent
notes is done, is the next time series:

\begin{figure}[H]
\begin{centering}
\includegraphics[scale=0.3]{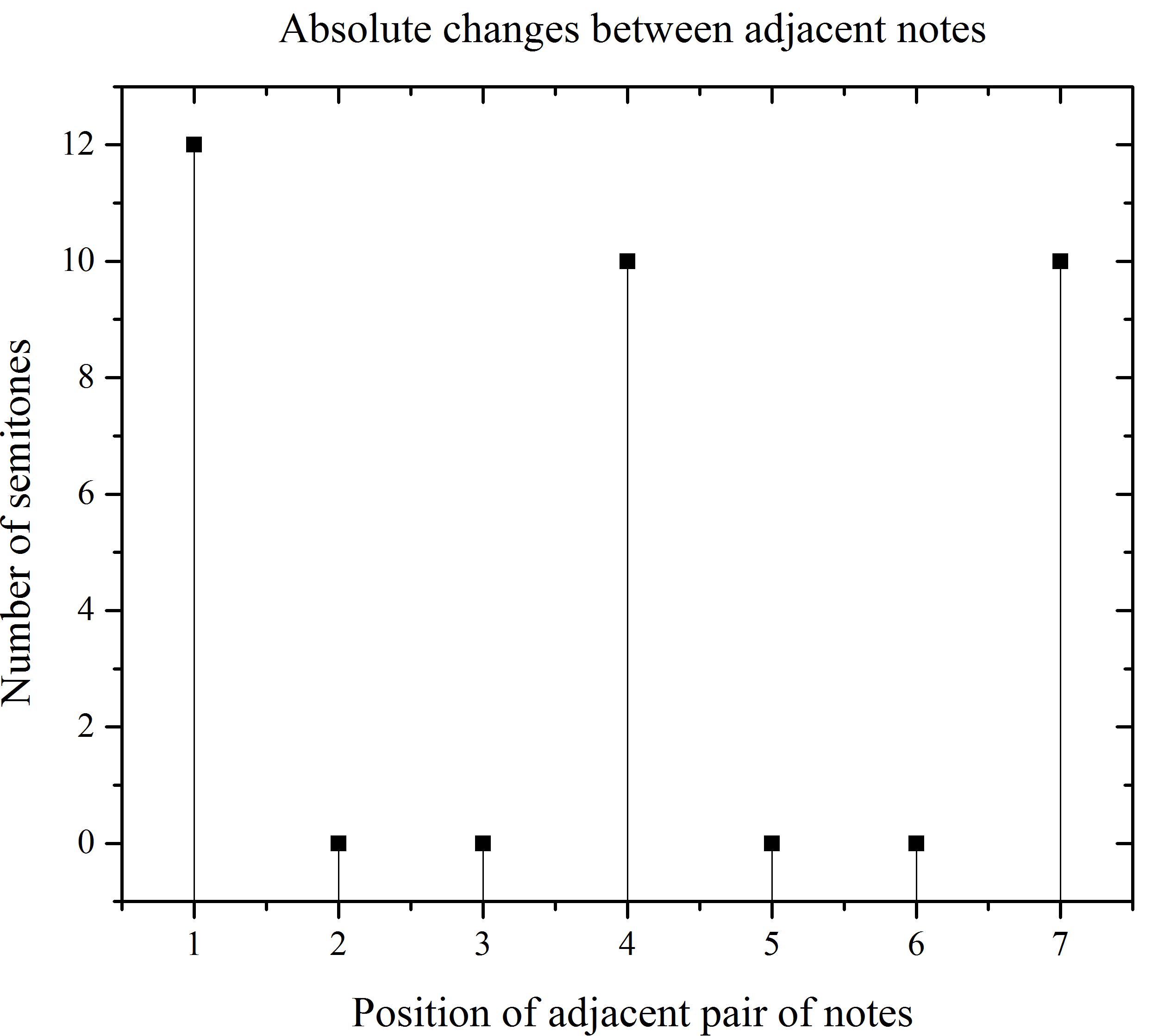}
\par\end{centering}

{\small{\caption{{\footnotesize{What we have in the pentagram is now reduced to a time
series, in wich the high is the diferences in semitones between notes.
For example if in figure \ref{pent} is used G clef, then the first
note is a G and the second one is also a G but an octave higer. Wich
means that between the first and the second notes there are 12 semitones
of diference. Between the second and the thirs there is no diference
so it apears as a cero, and so on.}}}
}}{\small \par}

\label{serie}
\end{figure}

\begin{figure}[H]
\begin{centering}
\includegraphics[scale=0.11]{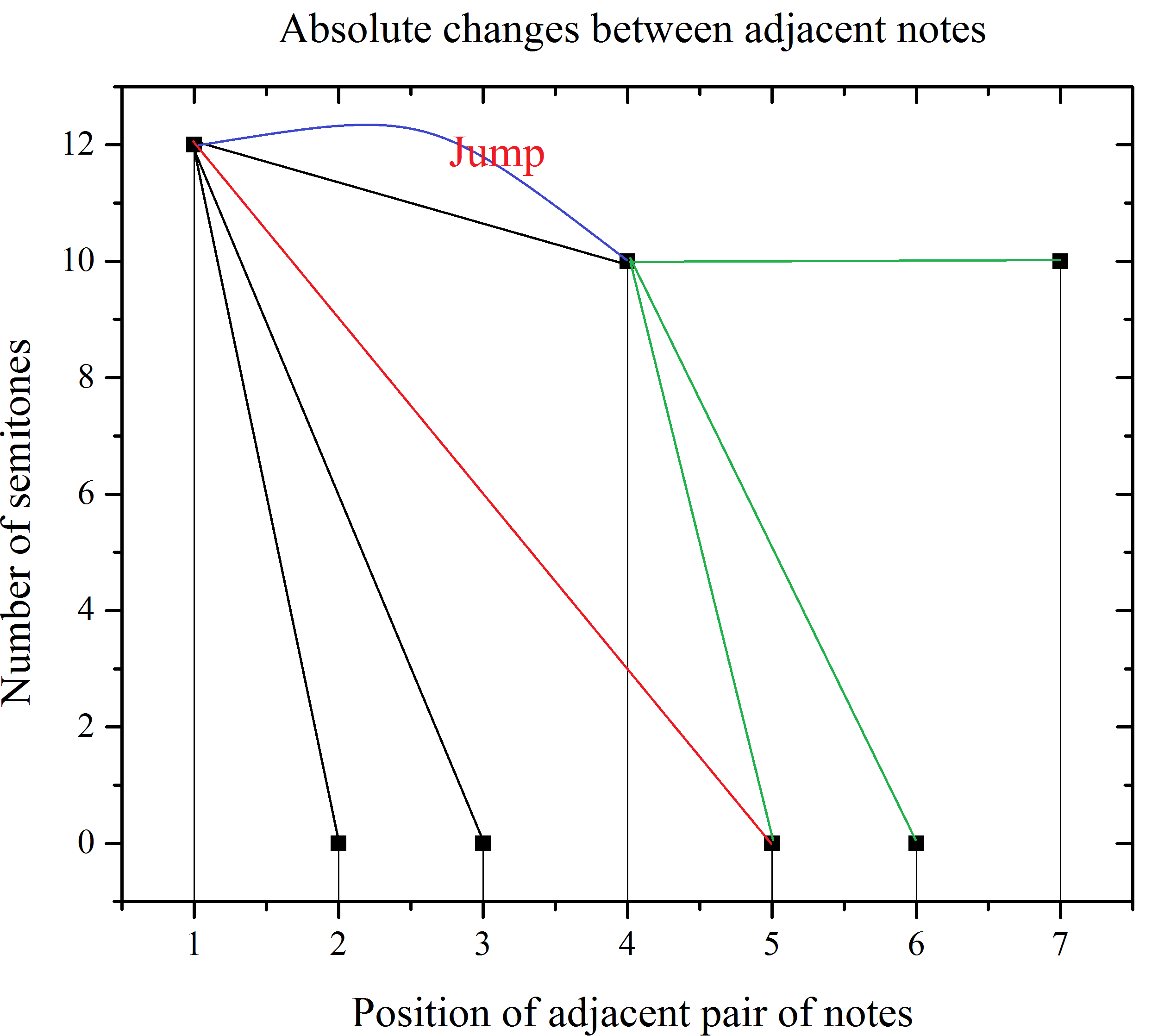}
\par\end{centering}

\label{brinco}\caption{{\footnotesize{We analyze figure \ref{serie} with visibility, we
changed that once visibility is cutted we stop and re-start one event
before. In this case we start in event number one and the visibility
is lost when we analyzed the first event in the series with number
5 so we re-start in event number 4, and we start visibility algorithm
again. Placing the algorithm one before the visibility is lost, is
to being placed in an event with higer visibility. }}}

\end{figure}

Once we have the rate of change series between adjacent semitones,
then we apply the visibility algorithm. In spite of using the visibility
rule we made some changes, instead of checking visibility between
point A and the following points in the series, we stop once the visibility
criteria isn't satisfy between A and some other point C in the time
series. In this point once we have stop the algorithm, a block is
formed, this block contain information of the semintones in the interval
{[}A,C), and the information is going to be codified using letters.
For example in figure \ref{brinco}, is the time series we are studying
under visibility. We start at event one in the series, we apply the
algorithm rules. Then we find that under this rules we are unable
to see event number five in the series, so we jump and restart the
algorithm in event number 4, always one before the visibility being
cutten. Once the the visibility criterion is not satisfy we do not
check for other events, this mean that we do not check if the visibility
criterion is true or false for event number six we simple move forward
in the time series.

We mencioned before that we are aimed to formed words, so when the
visibility is lost we create a block wich is going, from the event
in wich the algoritm has been started, to one before the visibility
has ben cutted. Neverteless in order to avoid confutions when reading
the blocks, we put a labels to the diference in semitones, and obtaining
a string of characters. Each caracter contain the information of the
diference in semitones, acording to the following table.

\begin{table}[H]
\begin{centering}
{\small{}}%
\begin{tabular}{cc}
\hline 
\multicolumn{1}{|c}{{\small{Nunmber of semitones }}} & \multicolumn{1}{c|}{{\small{Letter representation}}}\tabularnewline
\hline 
{\small{0}} & {\small{A}}\tabularnewline
{\small{1}} & {\small{B}}\tabularnewline
{\small{2}} & {\small{C}}\tabularnewline
{\small{3}} & {\small{D}}\tabularnewline
{\small{4}} & {\small{E}}\tabularnewline
$\cdot$ & $\cdot$\tabularnewline
$\cdot$ & $\cdot$\tabularnewline
$\cdot$ & $\cdot$\tabularnewline
\hline 
\end{tabular}
\par\end{centering}{\small \par}

{\small{\caption{{\footnotesize{Semitone Coding, using words instead of numbers in
order to generate blocks}}}
}}
\end{table}

With this, the information in figure \ref{serie} once is splitted
is going to be coded as the next blocks: MAAL-LAAL. Also is worth
it to mention, that if there is a difference above 25 semitones we
put a d lowercase, re-starting the letters with the $A$. For example
if there is a 26 diference in semitones then we put $Ad$. This doesn't
occur to often however it can occur. Also an importan thing is that
we consider a block, the union of two consecutives blocks. So the
final block above is going to be MAALAAL.

\section*{Results.}

We analyzed pieces from diferent composers like Bach, Beethoven, Chopin,
Scriabin, Satie among others. The majoritiy of ours files came from
kunstderfuge web site, nearly 1300 pieces were analyzed. First we
wanted to know if the difference series, transform as a scale free
network. In order to know this we run the visibility algorithm with
out any modification. What we saw was that the time series, was transformed
into a scale free network. This can be seeing in the next graphs,
we plot some of the reults found. The degree distribution follow a
power law on most cases and is aproximated to this behavior. 

\begin{figure}[H]
\begin{raggedright}
\includegraphics[scale=0.20]{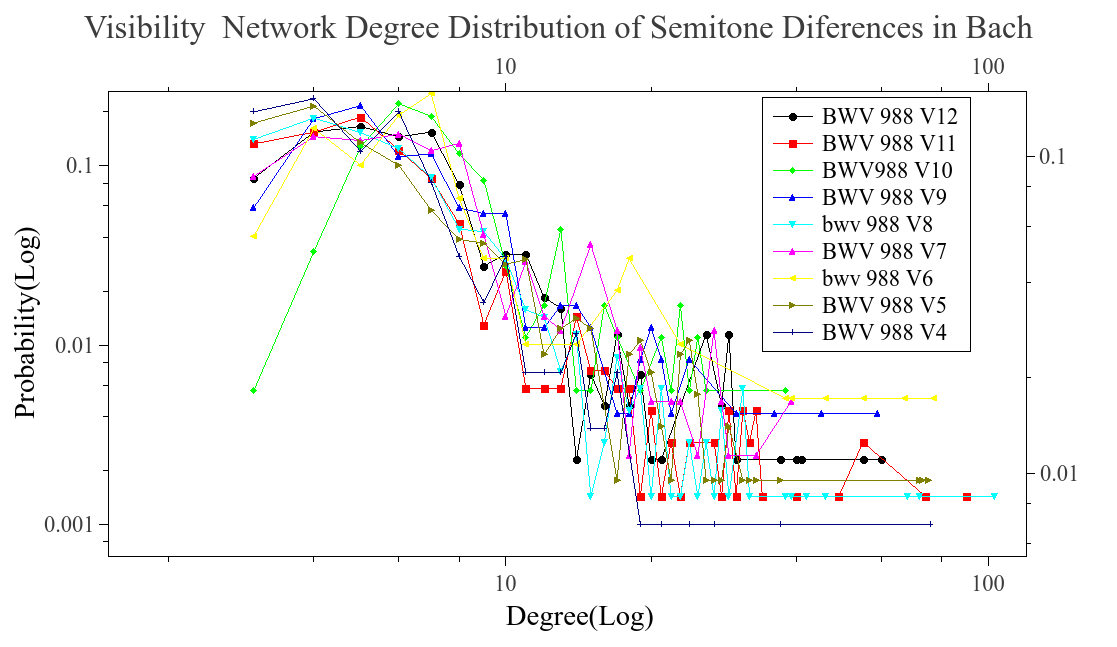}\includegraphics[scale=0.20]{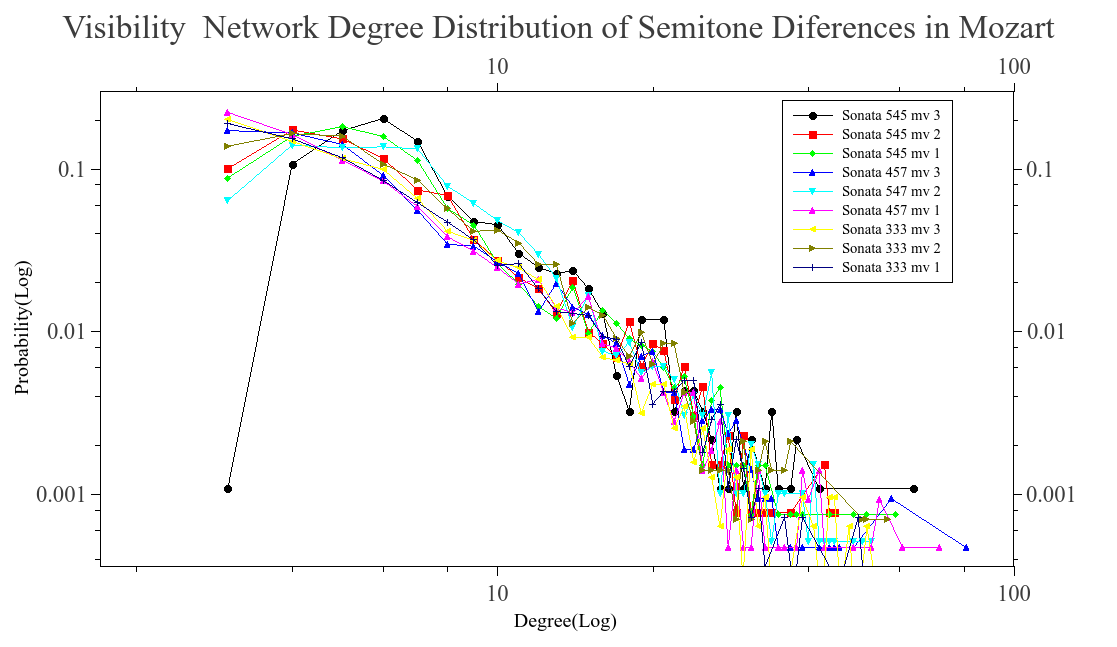} 
\par\end{raggedright}

\begin{raggedright}
\includegraphics[scale=0.20]{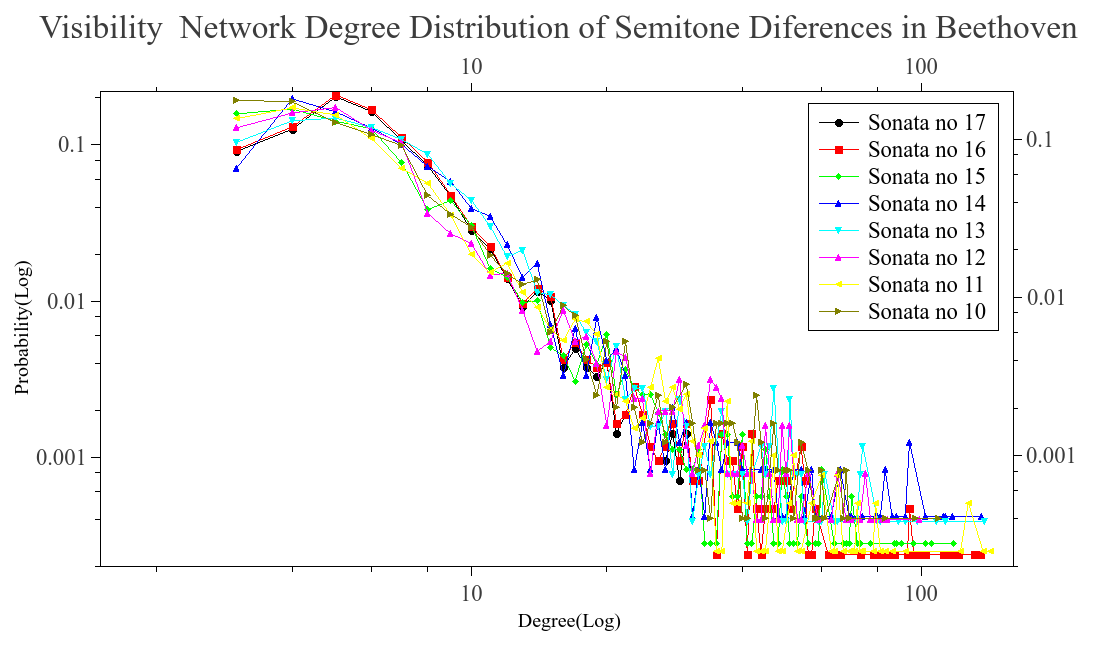}\includegraphics[scale=0.20]{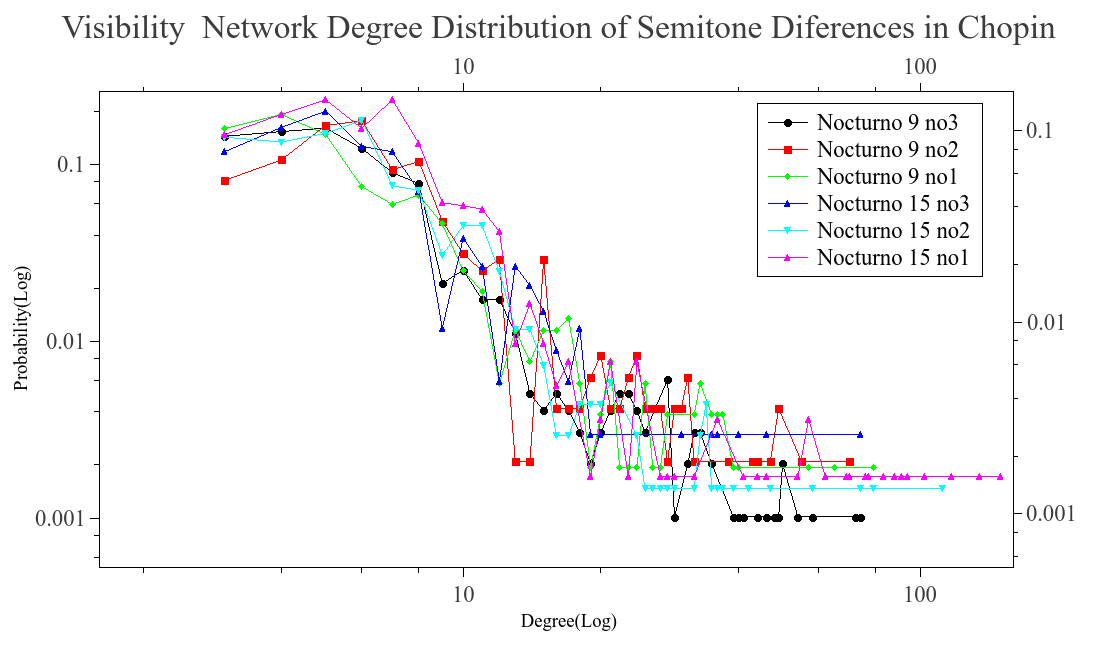}
\par\end{raggedright}
\begin{raggedright}
\includegraphics[scale=0.20]{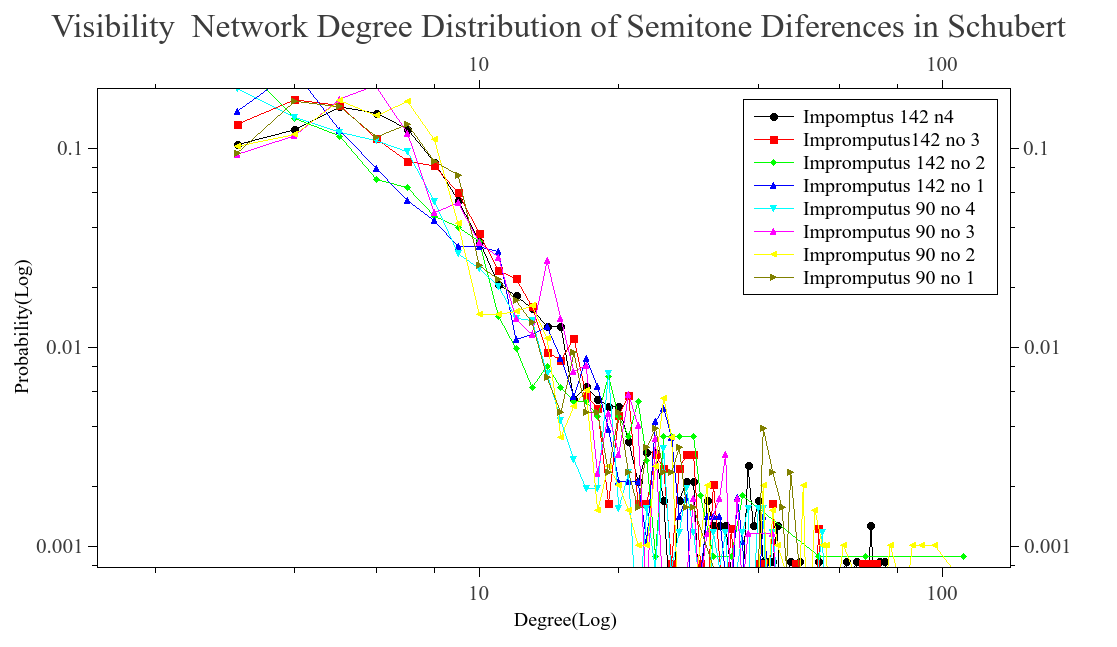}\includegraphics[scale=0.20]{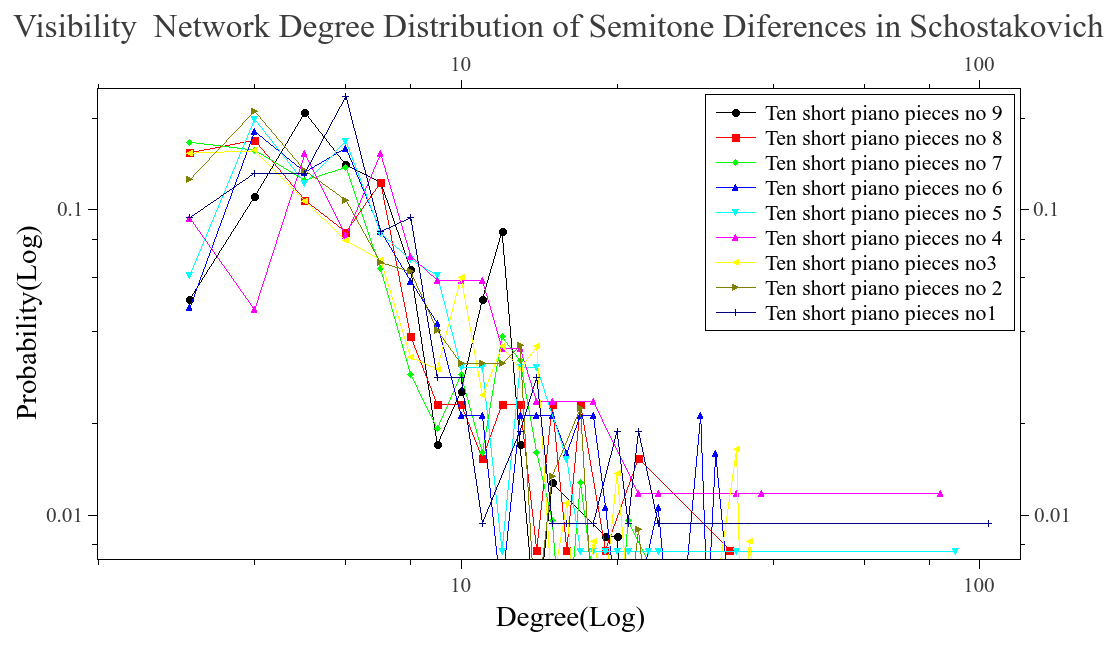}
\par\end{raggedright}

\caption{{\footnotesize{In the figures above we can see that the networks that
were formed by the visibility algorithm, follow a power law in the
degree distribution of the network.}}}

\end{figure}

Once we know that the networks are scale free. We take a subgroup
of the network, using the modifed algorithm. The blocks formed are
conteined in the network but posibly with less conections, becouse
of the modifications that were made when the visibility is lost. This
blocks is what we call musical words, and as we know, the lenguages
tent to follow Zipf's law. Thats why we wanted to know if the whole
network was scale free, as a maner to ensure that the words formed
also have this behavior and result in a Zipf's law when they are ranked.

In order to studie words, a data set was made. This data set contains
the words used by a certain author, and how often are used by this
composer. Thinking that every composer could be seen as a diferent
musical lenguage with his own words. Again we used the 1300 files,
and when we ploted, rank of the words used by the composer vs the
probability for a certain word, we find the following graph

\begin{figure}[H]
\begin{centering}
\includegraphics[scale=0.45]{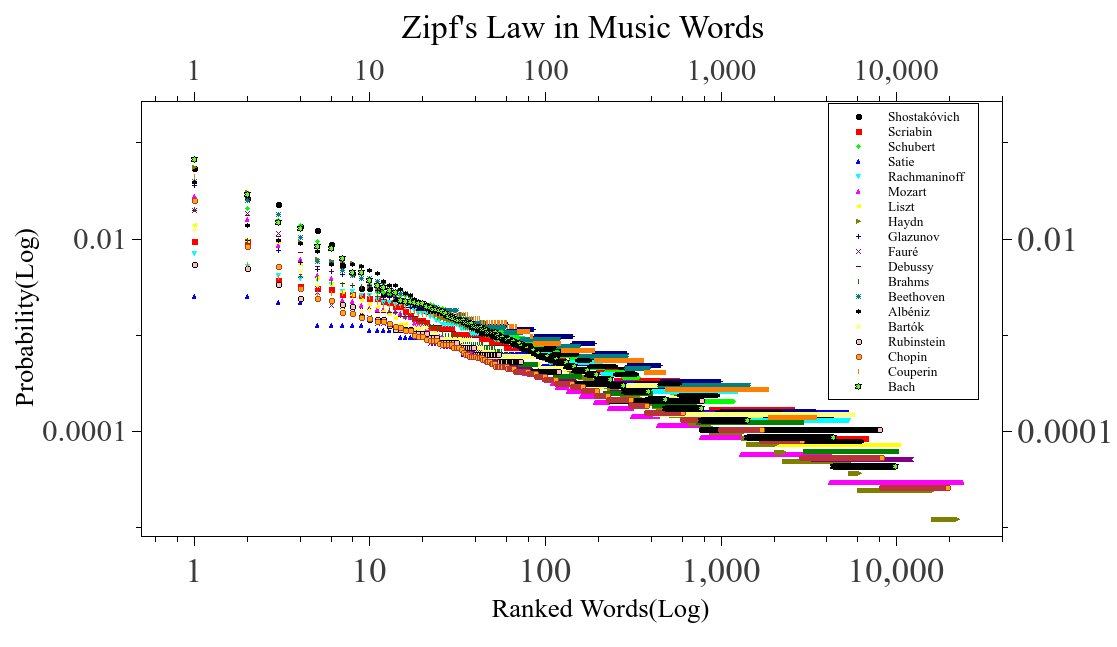}
\par\end{centering}

\caption{{\footnotesize{When the words formed are ranked and we do the graph
of ranked word vs probability of each word formed per compositor,
we can see that it tent to follow a power law distribution, meaning
that it has behavior in the majority, similar to Zipf 's law.}}}
\end{figure}

\begin{figure}[H]

\begin{centering}
\includegraphics[scale=0.35]{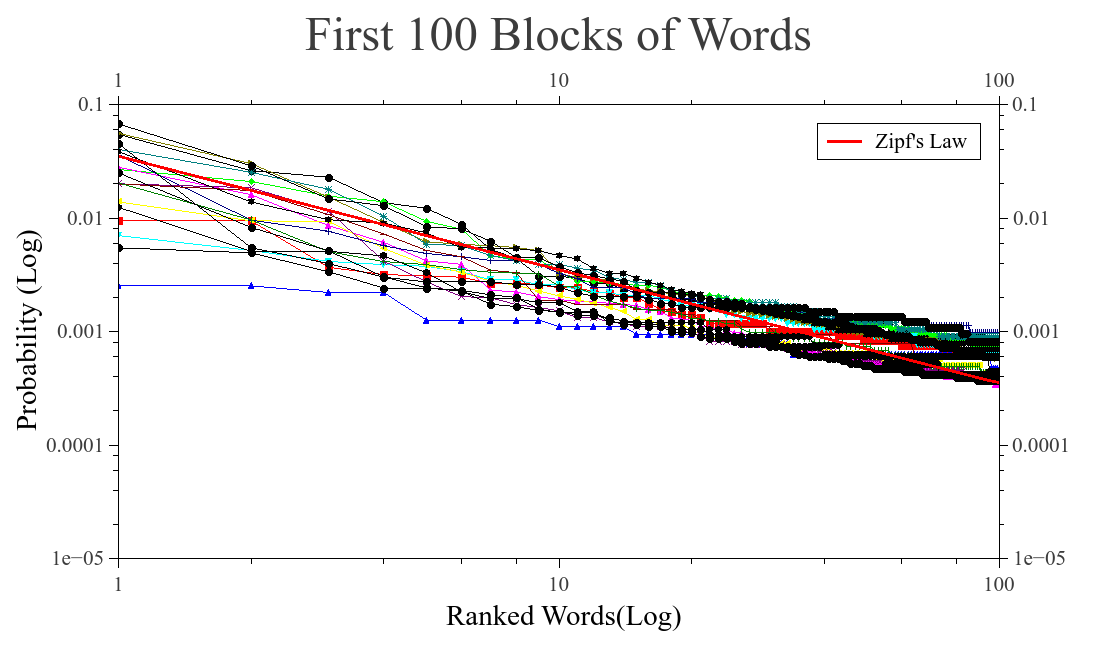}
\par\end{centering}

\caption{{\footnotesize{First 100 blocks}}}

\end{figure}

In the picture above we see that the blocks formed by the modified
algorithm tent to follow a Zipf's law, as lenguages does. With this
result, we can think that every composer generate his own lenguage.
The words in the lenguages are going to depend on the variations of
the nearest notes, remembering that every character in the block is
a mesure of the rate of change between adjacent notes in semitones.
Having the words of the diferent composers, seems rasonable to find
diferences among composers in the words formed. Meaning that we can
construct a time series based, on how a fixed word, is probable for
every composer. Thinking that every composer represent a diferent
time in the series.

\begin{minipage}[t]{1\columnwidth}%

\includegraphics[scale=0.3]{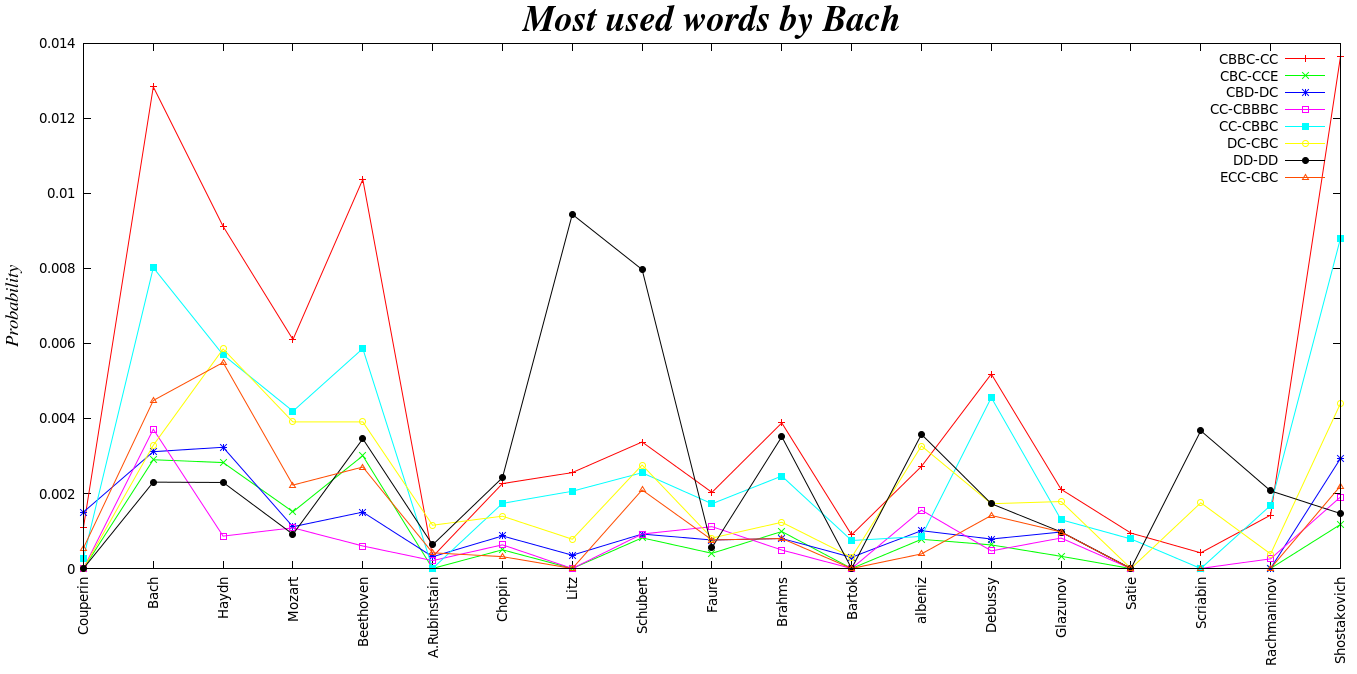}%
\end{minipage}

\begin{minipage}[t]{1\columnwidth}%
$\,$

$\,$

\includegraphics[scale=0.3]{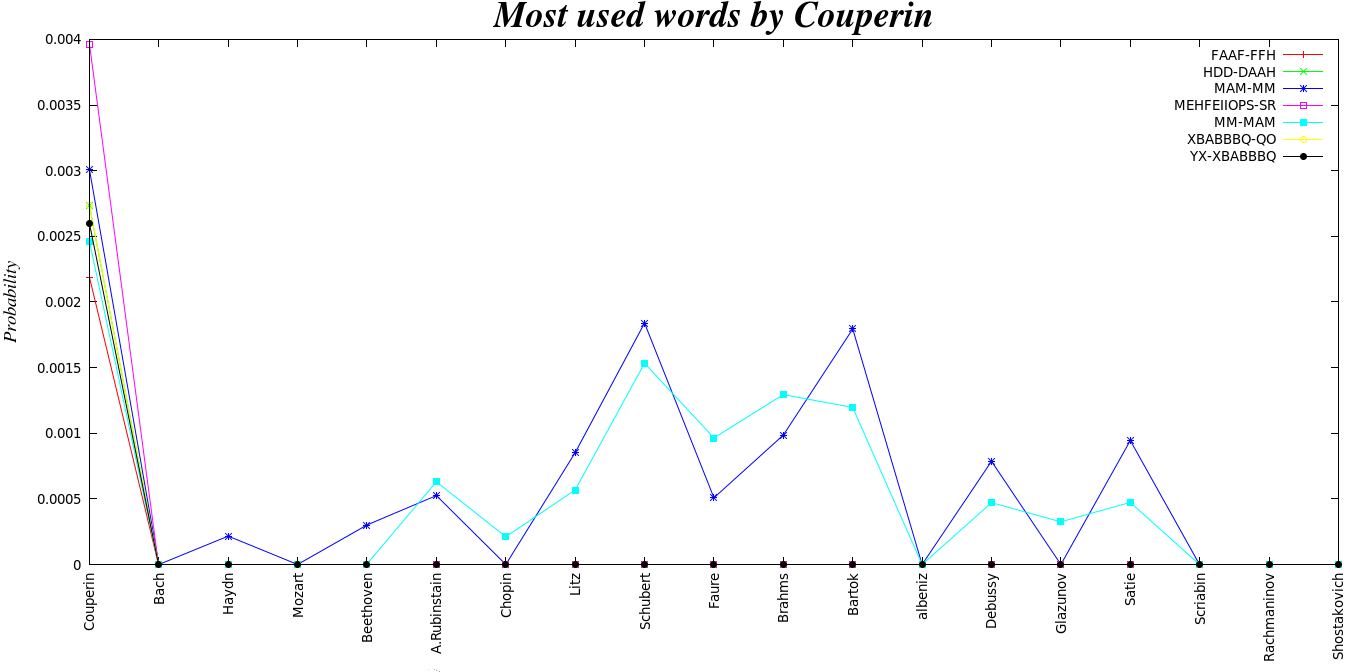}%
\end{minipage}

\begin{minipage}[t]{1\columnwidth}%
$\,$

$\,$

\includegraphics[scale=0.3]{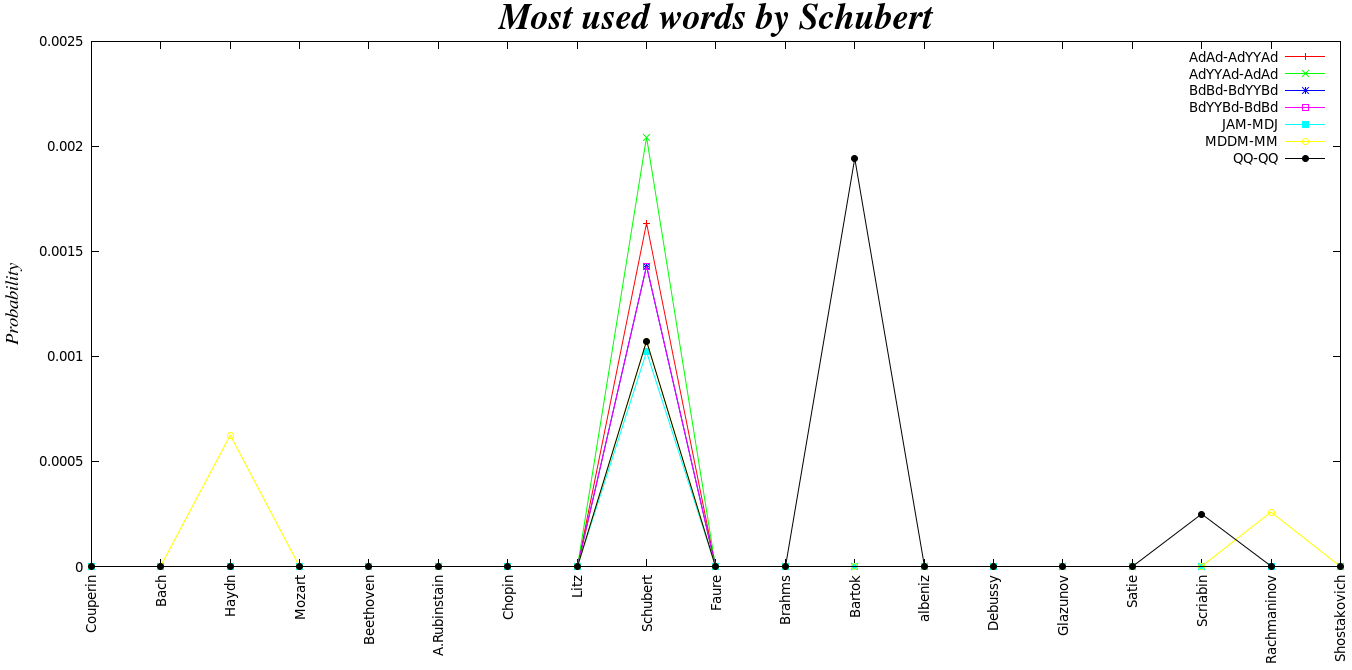}%
\end{minipage}

\begin{figure}[H]

\includegraphics[scale=0.3]{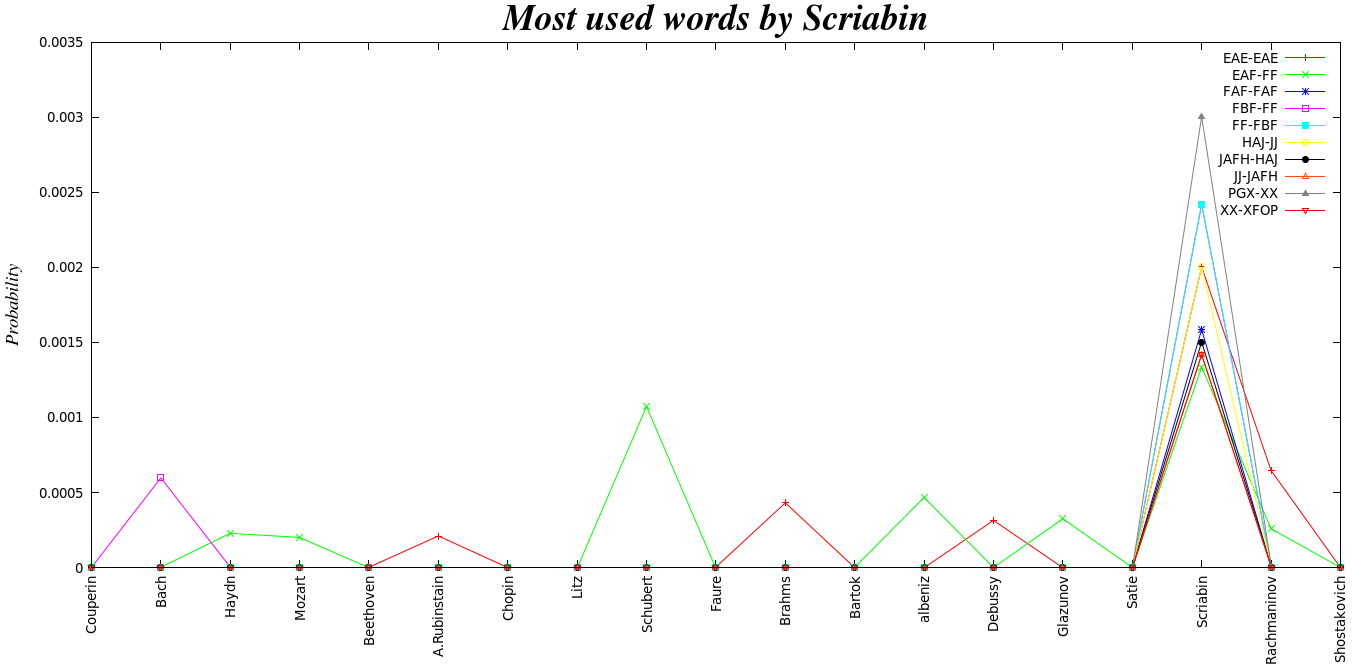}

\caption{{\footnotesize{Time series above are formed with the words we found.
We fixe a word that happen with a hig probability in a certain composer
and then this word is compared among the composers. We note that for
a fixed word, the probability change among composers it get higer
or lower. }}}

\end{figure}

As we see above the words provides us a way to see how change the
music from different times. For example, in early ages often we see
that the words begin whit an C or a D, and also is not so common to
find letters that are the last ones in the alphabet, among early composer.
So waht we can think is that in early ages of music the jump between
adjacent notes was really short in the majority of the times. Also
when we compare the most often words use by this early composer, we
can see that this words are used by more contemporany composers.The
exeption it could be Couperin, in wich we found words with higer alphabet
letters. When we see the graph of Couperin, we can see that most of
their words died soon or are not longer used, others are used latrer
on.

\begin{figure}[H]
\begin{centering}
\includegraphics[scale=0.5]{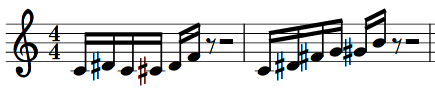}
\par\end{centering}

\caption{{\footnotesize{CBBC it is frequent in Bach. Nevertheless as we only
have information in adjacent notes we have multiple musical figures
that can match in CBBC}}}

\end{figure}

In contemporany composers we see that they like to introduce changes
in adyacent notes that are above an octave, this make a contrast with
early ages.Also among the most used words are some words that not
were used before or after. This don't ocure with early composers . 

\begin{figure}[H]
\begin{centering}
\includegraphics[scale=0.5]{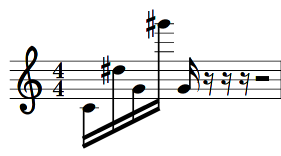}
\par\end{centering}

\caption{{\footnotesize{PGXX it is frequent in Scriabin. We can see that the
rate of change between adjacent nodes is different from Bach.}}}
\end{figure}

We can think that this most frequent words, are what makes the music
composers differents between them.

\end{document}